\documentclass[12pt]{article} 

\UseRawInputEncoding

\input{header.sty}

\usepackage[noblocks,affil-it]{authblk}

\usepackage{geometry} 
\usepackage{graphicx} 

\usepackage{enumitem}

\usepackage{booktabs} 
\usepackage{array} 
\usepackage{verbatim} 
\usepackage{subfig} 
\usepackage[linesnumbered]{algorithm2e}
\usepackage{lipsum} 

\usepackage{fancyhdr} 
\usepackage[nottoc,notlof,notlot]{tocbibind} 
\usepackage[titles,subfigure]{tocloft} 

\definecolor{Gray}{gray}{0.7}
\definecolor{lightGray}{gray}{0.8}

\newbox{\myorcidaffilbox}
\sbox{\myorcidaffilbox}{\large\includegraphics[height=1.7ex]{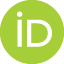}}
\newcommand{\orcid}[1]{\href{https://orcid.org/#1}{\usebox{\myorcidaffilbox}}}

\providecommand{\keywords}[1]{\textit{Keywords:} #1}

\title{\protect Joint Variable Selection of both Fixed and Random Effects for Gaussian Process-based Spatially Varying Coefficient Models}

\author[1,2]{Jakob A.\ Dambon\thanks{\emph{Corresponding author}. Email: \href{mailto:jakob.dambon@math.uzh.ch}{jakob.dambon@math.uzh.ch}, Address: Department of Mathematics, University of Zurich, Winterthurerstrasse 190, 8057 Zurich, Switzerland.}\orcid{0000-0001-5855-2017}}
\author[2]{Fabio Sigrist\thanks{Email: \href{mailto:fabio.sigrist@hslu.ch}{fabio.sigrist@hslu.ch}, Address: Institute of Financial Services Zug, Campus Zug-Rotkreuz, Suurstoffi 1, 6343 Rotkreuz, Switzerland.}\orcid{0000-0002-3994-2244}}
\author[1,3]{Reinhard Furrer\thanks{Email: \href{mailto:reinhard.furrer@math.uzh.ch}{reinhard.furrer@math.uzh.ch}, Address: Department of Mathematics, University of Zurich, Winterthurerstrasse 190, 8057 Zurich, Switzerland.
}\orcid{0000-0002-6319-2332}}
\affil[1]{Department of Mathematics, University of Zurich}
\affil[2]{Institute of Financial Services Zug, Lucerne University of Applied Sciences and Arts}
\affil[3]{Department of Computational Science, University of Zurich}

\begin{document}
\maketitle

\begin{abstract}
	Spatially varying coefficient (SVC) models are a type of regression model for spatial data where covariate effects vary over space. If there are several covariates, a natural question is which covariates have a spatially varying effect and which not. We present a new variable selection approach for Gaussian process-based SVC models. It relies on a penalized maximum likelihood estimation (PMLE) and allows variable selection both with respect to fixed effects and Gaussian process random effects. We validate our approach both in a simulation study as well as a real world data set. Our novel approach shows good selection performance in the simulation study. In the real data application, our proposed PMLE yields sparser SVC models and achieves a smaller information criterion than classical MLE. In a cross-validation applied on the real data, we show that sparser PML estimated SVC models are on par with ML estimated SVC models with respect to predictive performance. 
\end{abstract}

\keywords{adaptive LASSO, Bayesian optimization, coordinate descent algorithm, model-based optimization, penalized maximum likelihood estimation, spatial statistics}

\newpage

\section{Introduction}

Improved and inexpensive measuring devices lead to a substantial increase of spatial point data sets, where each observation is associated with (geographic) information on the observation location, say, a pair of latitude and longitude coordinates. Usually, these data sets not only consist of a large number of observations, but also include several covariates. Spatially varying coefficient (SVC) models are a type of regression models which offer a high degree of flexibility as well as interpretability for spatial data. In contrast to classical geostatistical models (see \citealp{Cressie2011}, for an overview), they allow the covariate's marginal effect, i.e., the corresponding coefficient, to vary over space. In \citet{JAD2020} a maximum likelihood estimation (\MLE) of Gaussian process (GP)-based SVC models is presented. In particular, the presented methodology not only scales well in the number of observations but also in the number of covariates. When modeling such data with SVC models, the large number of potential coefficients leads to the question: \emph{Which of the given covariates do indeed have a non-zero spatially varying coefficient?} This is similar to variable selection for classical fixed effects variables but, in addition, we also need to select random Gaussian process coefficients. While the literature on variable selection in general is extensive, it is very limited for SVC models. 

Variable selection for SVC models is usually done locally or globally over space. For instance, \citet{Smith2007} introduce a local selection procedure for SVC models on lattice data. \citet{Reich2010} proposed a Bayesian method that selects SVC globally, i.e., the coefficient enters the model equation either for all locations or for none. \citet{Vock2015} present a simultaneous local and global variable selection.  

There also exist non-model-based approaches for SVC selection, e.g., geographically weighted regression (GWR, \citealp{Fotheringham2002}). \citet{LiLam2018} present a local SVC approach selection based on the elastic net, while \citet{Lu2014} use an information criteria based global SVC selection for GWR. 

Concerning frequentist Gaussian process-based SVC models, existing variable selection methods which we list in the following are exclusively for the fixed effects parts. \citet{Huang2007} proposed a selection method between smoothing splines with deterministic spatial structure and kriging which is GP-based. The method relies on generalized degrees of freedom \citep{Ye1998} to assess the models. Another variable selection for the fixed effects was proposed by \citet{Wang2009}. Using penalized least squares, variable selection under a variety of penalty functions is conducted. In particular, the penalty function smoothly clipped absolute deviation (SCAD) suggested by \citet{FanLi2001} and used in the simulation study performs best for variable selection in spatial models. However, the error term of the spatial models is defined as strong mixing and in all of the numerical examples the sample size is very limited. \citet{ChuZhuWang2011} provide theoretical results for variable selection in a GP-based geostatistical model using a penalized maximum likelihood. Since MLE is computationally expensive, \citet{ChuZhuWang2011} also give results for variable selection under covariance tapering \citep{Furrer2006}.  Their penalized maximum likelihood estimation (\PMLE) procedure -- with and without covariance tapering -- is a one-step sparse estimation and therefore differs from \citet{FanLi2001}. The SCAD is the penalty function of choice in \citet{ChuZhuWang2011}, too.

In this article, we present a novel selection methodology for SVC models. The rest of the article is structured as follows. We introduce GP-based SVC models in Section~\ref{sec:SVCmodel}. In Section~\ref{sec:impl},  we propose our implementation of the selection problem. Numeric results of both a simulation study and a particular real data set application are given in Sections~\ref{sec:simu} and~\ref{sec:real}, respectively. In Section~\ref{sec:conc} we conclude.




\section{Variable Selection for GP-based SVC Models} \label{sec:SVCmodel}

\subsection{GP-based SVC Models}

Let $n$ be the number of observations, let $p$ be the number of covariates $\x^{(j)} \in \mathbb{R}^n, j = 1, ...,p$, for a fixed effect, and let $q$ be the number of covariates $\w^{(k)} \in \mathbb{R}^n, k = 1, ..., q$, with random coefficients, i.e., spatially varying coefficients. The fixed and random effects covariates do not necessarily need to be identical. Each observation $i$ is associated with an observation location $s_i$ in domain $D \subset \mathbb{R}^d, d\geq 1$. Further, let $\y \in \mathbb{R}^n$ be the vector of observed responses and $\bvarepsilon \sim \mathcal{N}_n(\0_n, \tau^2 \I_{n\times n})$ the error term (also called nugget). Without loss of generality, we assume the random coefficients to have zero mean and define a GP-based SVC model as
\begin{align}\label{eq:SVCmodelsingle}
	y_i = \sum_{j = 1}^p \mu_j x^{(j)}_i + \sum_{k = 1}^q  \eta_k(s_i) w^{(k)}_i + \varepsilon_i.
\end{align}
Here, the $k$th SVC is defined by a zero-mean GP $\eta_k(\cdot) \sim \mathcal{GP}(\0, c(\cdot, \cdot; \btheta_k))$ with covariance function $c$ and covariance parameters ${\btheta_k := (\rho_k, \sigma_k^2)}$, i.e., each SVC is parameterized by a range parameter $\rho_k$ and variance $\sigma_k^2$. We assume additional parameters like the smoothness of the covariance function $c$ to be known and that we can write
\begin{align}\label{eq:cov}
	c(s_l, s_m; \btheta_k) = \sigma_k^2 r\left(\frac{\| s_l - s_m \|_{A}}{\rho_k} \right),
\end{align}
where $r: \left[0, \infty \right) \rightarrow [ 0, 1]$ is a correlation function, $\| \cdot \|_A$ is an anisotropic geometric norm defined by a positive-definite matrix  $A \in \mathbb{R}^{d \times d}$ \citep{Wackernagel1995, Schmidt2020}. Note that \eqref{eq:cov} covers most of commonly used covariance functions such as the Mat\'ern class or the generalized Wendland class. For instance, the Mat\'ern covariance function is of the form of \eqref{eq:cov} with correlation function
 \begin{align}\label{eq:matern}
		r \left( u \right) = \frac{2^{1- \nu} }{\Gamma(\nu)} \left(\sqrt{2\nu} u \right)^\nu K_\nu \left(\sqrt{2\nu} u \right),
\end{align} 
where $\nu \in \mathbb{R}^{+}$ is the \emph{smoothness} parameter, $u = \| s_l - s_m \|_{A} / \rho$ is a scaled anisotropic distance, and $K_\nu$ is the modified Bessel function of the second kind and order $\nu$. For $\nu = 1/2$, equation~\eqref{eq:matern} reduces to the \emph{exponential function} $r(u) = \exp(-u)$ and the corresponding covariance function is therefore given by $c (s_l, s_m; \rho, \sigma^2 ) = \sigma^2 \exp(-\| s_l - s_m \|_{A}/\rho)$. Finally, we assume mutual prior independence between the SVCs $\eta_k(\cdot)$ as well as the nugget $\bvarepsilon$.

For the observed data, the GPs $\bfeta_k(\cdot)$ reduce to finite dimensional normal distributions with covariance matrices $\bSigma_k$ defined as $\left( \bSigma_k \right)_{l, m} := c(s_l, s_m; \btheta_k)$. Thus, we have $\bfeta_k(\s) \sim \mathcal{N}_n(\0_{n}, \bSigma_k)$ and under the assumption of mutual prior independence the joint effect is given by $\bfeta(\s)  \sim \mathcal{N}_{nq}(\0_{nq}, \bSigma_\bfeta)$ with $\bSigma_\bfeta := \diag (\bSigma_1, ..., \bSigma_q)\in \mathbb{R}^{nq \times nq}$. We denote two data matrices as $\X$ and $\W$, where we have $\X \in \mathbb{R}^{n \times p}$ such that the $j$th column is equal to $\x^{(j)}$ and $\W := \left( \diag (\w^{(1)}), ..., \diag (\w^{(q)}) \right) \in \mathbb{R}^{n \times nq}$. Let $\bmu := (\mu_1, ..., \mu_p)^\top$ and $\btheta := (\rho_1, \sigma_1^2, ..., \rho_q, \sigma_q^2, \tau^2)^\top$ be the unknown vectors of fixed effects and covariance parameters. Then the GP-based SVC model is given by
\begin{align}\label{eq:SVCmodel}
	\y = \X\bmu + \W\bfeta(\s) + \bvarepsilon
\end{align}
and is parametrized by
\begin{align}\label{eq:omega}
	\bomega := (\bmu^\top,\btheta^\top )^\top \in \Omega := \mathbb{R}^p \times \left(\mathbb{R}_{>0} \times \mathbb{R}_{\geq 0} \right)^q \times \mathbb{R}_{> 0} .
\end{align}
The response variable $\Y$ follows a multivariate normal distribution $\Y \sim \mathcal{N}_n\left( \X \bmu, \bSigma_\Y(\btheta) \right)$ with covariance matrix
\begin{align}
	\bSigma_\Y(\btheta) := \sum_{k = 1}^q \left(\w^{(k)} {\w^{(k)}}^{\top} \right) \odot \bSigma_k) + \tau^2 \I_{n\times n}
\end{align}
and has the following log-likelihood
\begin{align}\label{eq:LL}
	\ell(\bomega) = -\frac{1}{2} \left( n\log (2\pi) + 
		\log \det \bSigma_{\Y} (\btheta) + \left(\y - \X \bmu \right)^\top \bSigma_{\Y}(\btheta)^{-1}\left(\y - \X \bmu \right) \right).
\end{align}
\citet{JAD2020} provide a computationally efficient MLE approach to estimate $\bomega$ by maximizing \eqref{eq:LL}, which we denote as $\hat{\bomega}(\MLE) := \argmax_{\bomega \in \Omega} \ell(\bomega)$.

\subsection{Penalized Likelihood}

We introduce a penalized likelihood that will induce global variable selection for the GP-based SVC model \eqref{eq:SVCmodel}. Related to this, we say that for $\mu_j \neq 0$ an effect of $\x^{(j)}$ and for $\bfeta_k(\s) \neq \0_n$ an effect of $\w^{(k)}$ on the response is given.  Note that variable selection for the random coefficients corresponds to choosing between $\sigma_k^2>0$ and $\sigma_k^2=0$. For the special case that $\x^{(j)} = \w^{(j)}$ for some $j$, there are 3 possible cases for each covariate to enter a model \citep{Reich2010}:
\begin{enumerate}
	\item The $j$th covariate is associated with a non-zero mean SVC, i.e., there exist a non-zero fixed effect and a random coefficient. 
	\item There only exists a fixed effect $\mu_j \neq 0$ for the $j$th covariate, but $\bfeta_j(\s)$ is identical to 0. 
	\item The $j$th covariate enters the model solely through the zero-mean SVC, i.e., $\mu_j = 0$ and $\bfeta_j(\s)$ not identical to 0.
\end{enumerate}
The parameters within $\bomega$ that we penalize are therefore $\mu_j$ and $\sigma_k^2$. Given some penalty function $p(|\cdot |)$, we define ${p_\lambda(|\cdot|) := \lambda p(|\cdot|)}$, where $\lambda$ acts as a shrinkage parameter. We augment the likelihood function \eqref{eq:LL} with penalties for the mean and variance parameters. In general, each of the parameters $\mu_j$ and $\sigma_k^2$ have their corresponding shrinkage parameter $\lambda_{j}>0$ and $\lambda_{p+k}>0$, respectively, yielding the \emph{penalized log-likelihood}:
\begin{align}\label{eq:pLL}
	p\ell(\bomega) = \ell(\bomega) - n\sum_{j = 1}^{p} p_{\lambda_j}(|\mu_j|) - n\sum_{k = 1}^{q} p_{\lambda_{p+k}}(|\sigma_k^2|).
\end{align}
For a given set of shrinkage parameters $\lambda_j, 1 \leq j \leq p + q$, we maximize the penalized likelihood function to obtain a \emph{penalized maximum likelihood estimate}:
\begin{align}\label{eq:PMLE}
	\hat{\bomega}(\PMLE) := \argmax_{\bomega \in \Omega} p\ell(\bomega ).
\end{align}
Note the similarity to \citet{Bondell2010} and \citet{Ibrahim2011} who present a joint variable selection by individually penalizing the fixed and random effects in linear mixed effects models. \citet{Mueller2013} give an overview of such selection and shrinkage methods.

Finally, one usually assumes the penalty function to be singular at the origin and non-concave on $(0, \infty)$ to ensure that $\hat{\bomega}(\PMLE)$ has favorable properties such as sparsity, continuity, and unbiasedness \citep{FanLi2001}. From now on, we will consider the $L_1$ penalty function, i.e., $p(| \cdot |) = | \cdot |$ \citep{Tibshirani1996}. 


\section{Penalized Likelihood Optimization and Choice of Shrinkage Parameter}\label{sec:impl}

In this section, we show how to find the maximizer of the penalized log-likelihood and how we choose the tuning parameters $\lambda_j$.

\subsection{Optimization of Penalized Likelihood}

In the following, we show how a maximizer of the penalized log-likelihood \eqref{eq:pLL} can be found for given $\lambda_j$. This is equivalent to minimizing the negative penalized log-likelihood. We propose a (block) coordinate descent (CD) which cyclically optimizes over the mean parameters $\bmu$ and covariance parameters $\btheta$, i.e., 
\begin{align}
	\bmu^{(t+1)} &:= \argmin_{\bmu\in \mathbb{R}^p} \left[ -p\ell(\bmu| \btheta^{(t)}) \right], \label{eq:CDAmu} \\
	\btheta^{(t+1)} &:= \argmin_{\btheta \in \Theta} \left[ -p\ell(\btheta | \bmu^{(t+1)}) \right], \label{eq:CDAtheta}
\end{align}
for $t \geq 0$. The initial values are given by the ML estimate $\btheta^{(0)}:=\hat{\btheta}(\MLE)$. Algorithm~\ref{alg:CDA} summarizes this block coordinate descent approach. In the following, we show how the component wise minimizations \eqref{eq:CDAmu} and \eqref{eq:CDAtheta} are realized. 

\begin{algorithm}
	\KwData{shrinkage parameters $\lambda_j$, convergence threshold $\delta$, ML estimate $\hat{\btheta}(\MLE)$, objective function $p\ell(\cdot)$}
 	\KwResult{$\hat{\bomega}(\PMLE)$}
 	Initialize $t \leftarrow 0, \btheta^{(0)} \leftarrow \hat{\btheta}(\MLE)$ \;
 	\Repeat{$\|\btheta^{(t)}-\btheta^{(t-1)}\|_1 / \|\btheta^{(t-1)}\|_1 < \delta$}{
  		$\bmu^{(t+1)} \leftarrow \argmin_{\bmu \in \mathbb{R}^p} \left[ -p\ell(\bmu|  \btheta^{(t)}) \right]$\; \label{alg:CDAeq1}
  		$\btheta^{(t+1)} \leftarrow \argmin_{\btheta\in \Theta}  \left[ -p\ell(\btheta| \bmu^{(t+1)}) \right]$\;\label{alg:CDAeq2}
  		$t \leftarrow t + 1$ \;
  	}
  	$\hat{\bomega}(\PMLE) \leftarrow \bigl( {\btheta^{(t)}}^\top, {\bmu^{(t)}}^\top \bigr)^\top$\;
 \caption{General coordinate descent algorithm for penalized likelihood.}\label{alg:CDA}
\end{algorithm}

\subsubsection{Optimization over Mean Parameters}

We fix the covariance parameters $\btheta^{(t)}$ for some $t \geq 0$. The optimization step for the mean parameter (line~\ref{alg:CDAeq1}, Algorithm~\ref{alg:CDA}) simplifies to 
\begin{align*}
	\bmu^{(t+1)} &= \argmin_{\bmu \in \mathbb{R}^p} \left[ -\ell(\bmu | \btheta^{(t)}) + n\sum_{j = 1}^{p} \lambda_j |\mu_j| \right]\\
	&= \argmin_{\bmu \in \mathbb{R}^p} \left[ \frac{1}{2}\left(\y - \X \bmu \right)^\top \bSigma_{\Y}(\btheta^{(t)})^{-1}\left(\y - \X \bmu \right) + n\sum_{j = 1}^{p} \lambda_j |\mu_j| \right]
\end{align*}
We note that the first term is a generalized least square estimate for a linear model with $\Y \sim\mathcal{N}_n \left(\X\bmu, \bSigma_{\Y}(\btheta^{(t)})\right)$. Using the Cholesky decomposition $\L$ of $\bSigma_{\Y}(\btheta^{(t)})$ and a simple variable transformation $\tilde{\y} := \L^{-1} \y$ and $\tilde{\X} := \L^{-1} \X$, the objective function simplifies to
\begin{align*}
	\bmu^{(t)} = \argmin_{\bmu \in \mathbb{R}^p} \left[ \frac{1}{2n}\|\tilde{\y} - \tilde{\X}\bmu \|_2^2 + \sum_{j = 1}^{p} \lambda_j |\mu_j| \right].
\end{align*}
We note that this objective function coincides with a LASSO for a classical linear regression model \citep{Tibshirani1996} with individual shrinkage parameters per coefficient.

\subsubsection{Optimization over Covariance Parameters}

We optimize over the covariance parameters $\btheta$ with fixed mean parameter $\bmu^{(t+1)}$ (c.f. Algorithm~\ref{alg:CDA}, line~\ref{alg:CDAeq2}). We rearrange this optimization problem writing $\btheta^{(t+1)} = \argmin_{\btheta\in \Theta} f(\btheta)$ with the following objective function:
\begin{align} \label{eq:objfun}
	f (\btheta) := -\ell \left( \bigl({\bmu^{(t)}}^\top, \btheta^\top \bigr)^\top \right) +n \sum_{k = 1}^{q} \lambda_{p + k} |\sigma^2_k|.
\end{align}
We use numeric optimization to obtain $\btheta^{(t+1)}$. In particular, we use a quasi Newton method \citep{Byrd1995} for the numeric optimization. Note that for a $\kappa \in \{ 1, ..., q \}$ with $\sigma_\kappa^2 = 0$, the corresponding range $\rho_\kappa$ is not identifiable and might induce non-convergence issues. The following proposition ensures that in case described above the optimization is well behaved.

\begin{prop}\label{prop:gr0}
	Let $r$ be a correlation function for a GP-based SVC model as given above. Assume that the derivative of $r$ exists and is bounded, i.e., $|r'|<C$ for some constant $C \geq 0$. Let $B_\kappa:= \{\btheta \in \Theta : \sigma_\kappa^2 = 0\}$ for $\kappa \in \{1, ..., k\}$. Then the objective function $f$ defined in \eqref{eq:objfun} fulfills
	\begin{align*}
		\frac{\partial}{\partial \rho_\kappa} f(\b_\kappa) = 0,
	\end{align*}
	for all $\b_\kappa \in B_\kappa$ and $\kappa \in \{1, ..., k\}$.
\end{prop}

\begin{proof}
	The proof is given in the Appendix~\ref{proof:gr0}.
\end{proof}

The additional assumption $|r'|<C$ is quite weak and fulfilled by most of covariance functions used for modeling GP. The partial derivative with respect to $\rho_\kappa$ is 0 and therefore the approximation of the gradient function for $\rho_\kappa$ yields values very close or identically to 0. Therefore, in the case of $\sigma_\kappa^2 = 0$, the numeric optimization will stop making any adjustments along the $\rho_\kappa$ direction. 

\subsection{Choice of Shrinkage Parameters}

We use the adaptive LASSO (ALASSO, \citealp{Zou2006}) for the parameters $\lambda_j$ given by
\begin{align}\label{eq:adappenals}
	\lambda_j := \frac{\lambda_\bmu}{|\hat{\mu}_j|}, \quad \quad \lambda_{p + k} := \frac{\lambda_\btheta}{|\hat{\sigma}^2_k|},
\end{align}
where we use the ML estimates $\hat{\mu}_j(\MLE)$ and $\hat{\sigma}^2_k(\MLE)$ to weight the shrinkage parameters $(\lambda_\bmu, \lambda_\btheta) \in \Lambda := \left(\mathbb{R}_{>0}\right)^2$. These two parameters account for the differences between the mean and variance parameters. This leaves us with the task to find a sensible choice of shrinkage parameters $\blambda := (\lambda_\bmu, \lambda_\btheta) \in \Lambda$. We choose these parameters by maximizing an information criterion (IC) which combines the goodness of fit (GoF) and model complexity (MC), i.e, 
\begin{align*}
	\textnormal{IC} = \textnormal{GoF} + \textnormal{MC}.
\end{align*}
An alternative approach that is computationally more expensive is to use cross-validation. To emphasize the dependency on $\blambda$, we introduce the short hand notation $\hat{\bomega}_\blambda$ for denoting the PML estimate of $\bomega$ for a given $\blambda$. The corresponding mean parameters and variances are given by $\hat{\bmu}_\blambda$ and $\hat{\bsigma}^2_\blambda$, respectively. We use a BIC type information criterion which is given by
\begin{align}\label{eq:BIC}
	\BIC = -2\ell(\hat{\bomega}_\blambda) + \log(n)\left(\|\hat{\bmu}_\blambda \|_0 + \| \hat{\bsigma}^2_\blambda\|_0 \right),
\end{align}
where $\| \cdot \|_0$ is the count of non-zero entries. That is, the MC is captured via the number of non-zero fixed effects and non-constant random coefficients. Of course, there exist various ICs like, for instance, the corrected Akaike IC (cAIC, see \citealp{Mueller2013}, for an overview). In our empirical experience the BIC performs best with respect to variable selection which is in line with \citet{Schelldorfer2011}.

Our goal is to find a minimizer $\blambda$, i.e., ${\hat{\blambda}_\BIC := \argmin_{\blambda \in \Lambda}\BIC(\blambda)}$. A single evaluation of $\BIC(\blambda)$ is computationally expensive, which is why we want the number of evaluations as low as possible. To this end, we use model-based optimization (MBO, also known as Bayesian optimization, \citealp{Jones2001, Koch2012, Horn2016}). For our objective function $\BIC(\blambda)$, we initialize the optimization by drawing a Latin square sample of size $n_{init}$ from $\Lambda$ which we define as $\{ \blambda^{(1)}, ..., \blambda^{(n_{init})} \}  \subset \Lambda$. Then we compute $\xi^{(i)} = \BIC(\blambda^{(i)})$ to obtain the tuples $\bigl(\xi^{(i)}, \blambda^{(i)}\bigr)$. Now we can fit a surrogate model to the tuples. In our case, we assume a kriging model defined as a Gaussian process with a Mat\'ern covariance function of smoothness $\nu = 3/2$, c.f. equation~\eqref{eq:matern}. That is, given the observed tuples, we have $\Xi(\blambda)\sim \mathcal{N} \left(\hat{\mu}(\blambda), \hat{s}^2(\blambda) \right)$ as our surrogate model. An infill criterion is used to find the next, most promising $\blambda^{(n_{init} + 1)}$. We use the expected improvement (EI) infill criterion which is derived from the current surrogate model $\Xi(\blambda)$. The EI at $\blambda$ is defined as
\begin{align}\label{eq:EI}
	\textnormal{EI}(\blambda) = E_{\Xi}\left(\max \{ \xi_{min} - \Xi(\blambda), 0 \} \right),
\end{align}
where $\xi_{min}$ denotes the current BIC minimum. In the case of a GP-based surrogate model, the EI \eqref{eq:EI} can be expressed analytically. A separate optimization of \eqref{eq:EI} returns the best parameter $\blambda^{(n_{init} + 1)}$ according to the infill criteria and then the BIC is calculated. The tuple $\left(\xi^{(n_{init} + 1)}, \blambda^{(n_{init} + 1)}\right)$ is added to the existing tuples and surrogate model is updated. This procedure is repeated $n_{iter}$ times. Finally, the minimizer of the BIC from the set $\{\blambda^{(i)} : i = 1, ..., n_{init} + n_{iter}\}$ is returned. For more details regarding MBO refer to \citet{R:mlrMBO}.

\subsection{Software Implementation}

The described \PMLE\ approach is implemented in the \R\ package \pkg{varycoef} \citep{R:varycoef}. It contains the discussed MBO which is implemented using the \R\ package \pkg{mlrMBO} \citep{R:mlrMBO} as well as a grid search to find the best shrinkage parameters. Further, the \pkg{varycoef} package implements our proposed optimization of the penalized likelihood using a coordinate descent. The optimization over the mean parameters is executed using a classical adaptive LASSO with the \R\ package \pkg{glmnet} \citep{Friedman2010}. For the optimization over the covariance parameters, we use the quasi Newton method \texttt{"L-BFGS-B"} \citep{Byrd1995} which is available in a parallel version in the \R\ package \pkg{optimParallel} \citep{R:optimParallel}.

\section{Simulation}\label{sec:simu}

\subsection{Set-Up}

We simulate $N = 100$ data sets consisting of $n = 15^2 = 225$ samples with observation locations $\s_i, i = 1, ..., n,$ from a $15 \times 15$ \emph{perturbed grid}, c.f. Appendix~\ref{app:per_grid} and \citet{Furrer2016, JAD2020}. As in \citet{Tibshirani1996} and his suggested simulation set up, which has been assumed (e.g.~\citealp{FanLi2001}) or adapted (e.g.~\citealp{Li2008, Ibrahim2011}) in other simulation studies for variable selection, we use $p=q=8$ covariates which are sampled from a multivariate zero mean normal distribution such that the covariance between $\x^{(j)}$ and $\x^{(k)}$ is $\gamma^{|j-k|}$ with $\gamma = 0.5$. For each simulation run, we sample the covariates $\x^{(j)}$ which are both associated with a fixed and random effect. Let $\X$ and $\W$ be the corresponding data matrices as defined above. Concerning the fixed effects parameters, we assume
\begin{align}\label{eq:true_mu}
	\bmu = (3, 1.5, 0, 0, 2, 0, 1, 0)^\top,
\end{align}
in order to obtain a balanced design in the sense that for each possible combination of zero and non-zero parameters of both the fixed and random effects there are two cases. Let $\bfeta_k(\s) \sim \mathcal{N}_n(\0_n, \bSigma_k)$ for $k = 1, ..., q$ be the GP-based SVCs defined by an exponential covariance function with corresponding parameters provided in Table~\ref{tab:simcovpars}. For $k\in \{2, 4, 7, 8\}$ the respective variance $\sigma_k^2$ is zero and therefore the respective true SVCs $\bfeta_k(\s)$ are constant zero. Hence, the true model contains two of each covariates with a non-zero mean SVC ($k = 1, 5$), constant non-zero mean effects ($k = 2, 7$), zero mean SVC ($k = 3, 6$), and without any effect ($k = 4, 8$). Adding a sampled nugget effect $\bvarepsilon \sim \mathcal{N}_n(\0_n, \tau^2\I_{n\times n})$ with nugget variance $\tau^2 = 0.1$ and independent of the GPs we can compute the response $\y$.

We compare methodologies for estimating the SVC model, c.f. \eqref{eq:SVCmodel}, 
or the oracle SVC model
\begin{align}\label{eq:oracleSVC}
	\y &= \X  (\mu_1, \mu_2, 0, 0, \mu_5, 0, \mu_7, 0)^\top + \sum_{k \in  \{1, 3, 5, 6\}} \bfeta_k(\s) \odot \x^{(k)} + \bvarepsilon.
\end{align}
The latter one is called the oracle model as it assumes the true data generating covariates to be known and excludes all other parameters from their respective estimation. As a reference, we will use two classical MLE approaches without any shrinkage or variable selection to estimate \eqref{eq:SVCmodel} and \eqref{eq:oracleSVC} labeled \textsf{MLE} and \textsf{Oracle}, respectively. The third methodology is our novel approach which estimates \eqref{eq:SVCmodel} and we denote it by \textsf{PMLE}. For the coordinate descent algorithm, we set a relative convergence threshold of $\delta = 10^{-6}$ and a maximum of $T_{max} = 20$ iterations. Note that this upper limit on the number of iterations was never attained in our simulations. Further, the range of shrinkage parameters was set to $(10^{-6}, 1)$, i.e., $\blambda = (\lambda_\bmu, \lambda_\btheta)^\top \in (10^{-6}, 1) \times (10^{-6}, 1)$. Finally, the shrinkage parameters $\hat{\blambda}_{\textnormal{BIC}}$ are estimated by MBO using $n_{init} = 10$ initial evaluations and $n_{iter} = 10$ iteration steps. The infill criterion is the previously defined expected improvement given in equation \eqref{eq:EI}. The shrinkage parameter ranges as well as the number of initial evaluations and iteration steps was chosen such that we obtain reasonable results under a feasible time and computational resource constraint. The methods \textsf{PMLE}, \textsf{MLE}, and \textsf{Oracle} are implemented using \pkg{varycoef} \citep{R:varycoef}. Further details are given in Appendix~\ref{app:num_opt}.

\begin{table}[ht]
\centering
\caption{True GP parametrization for simulation study. The ranges of zero variance GP are not identifiable which we denote by ``--". To ease readability, parameters identical to 0 are given by a sole zero, i.e., \emph{not} by ``0.000".} 
\label{tab:simcovpars}
\begin{tabular}{lrrrrrrrr}
  \hline \textbf{Parameters} & \multicolumn{8}{c}{\textbf{True GP Covariance Paramaters}, $k = $}  \\  \cmidrule(lr){2-9}  & 1 & 2 & 3 & 4 & 5 & 6 & 7 & 8 \\ 
  \hline
Variance $\sigma_k^2$ & 0.200 & \phantom{0.00}0 & 0.250 & \phantom{0.00}0 & 0.250 & 0.200 & \phantom{0.00}0 & \phantom{0.00}0 \\ 
  Range $\rho_k$ & 0.200 & -- & 0.100 & -- & 0.075 & 0.100 & -- & -- \\ 
   \hline
\end{tabular}
\end{table}

\subsection{Results}

\begin{figure}
\begin{center}
\includegraphics[width = \textwidth]{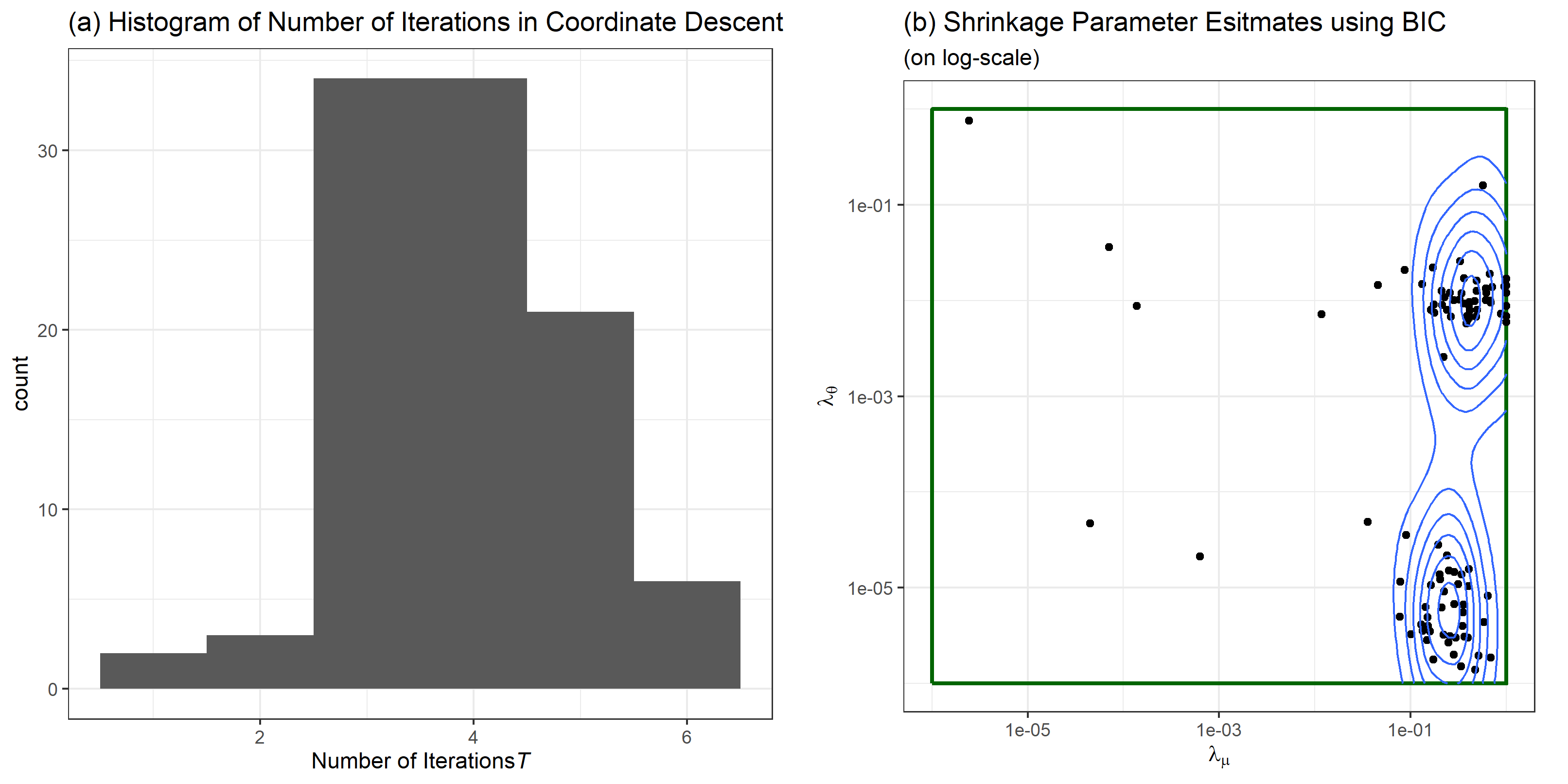}
\caption{The coordinate descent with (a) the number of iterations steps $T$ and (b) the MBO estimated $\hat{\blambda}$ under the BIC.} \label{fig:CD_BIC}
\end{center}
\end{figure}

First, we check the convergence properties of the CD over all simulations. This is depicted in Figure~\ref{fig:CD_BIC}. Facet (a) shows a histogram of the number of evaluated iterations $T$, where the median over all iterations was $4$. This shows that the CD algorithm converges quickly. Second, the estimated shrinkage parameters are depicted in Figure~\ref{fig:CD_BIC}(b). We observe that there exist two regimes with respect to $\lambda_\btheta$. The borders of the shrinkage parameter space used in the MBO are depicted darkgreen. There are some estimates very close to the border. Upon closer inspection of actual parameter estimates (see below) we could not find any substantial difference between parameter estimates obtained by $\hat{\blambda}_{\textnormal{BIC}}$ close to the boundary compared to all other $\hat{\blambda}_{\textnormal{BIC}}$. 

\begin{figure}
\begin{center}
\includegraphics[width = \textwidth]{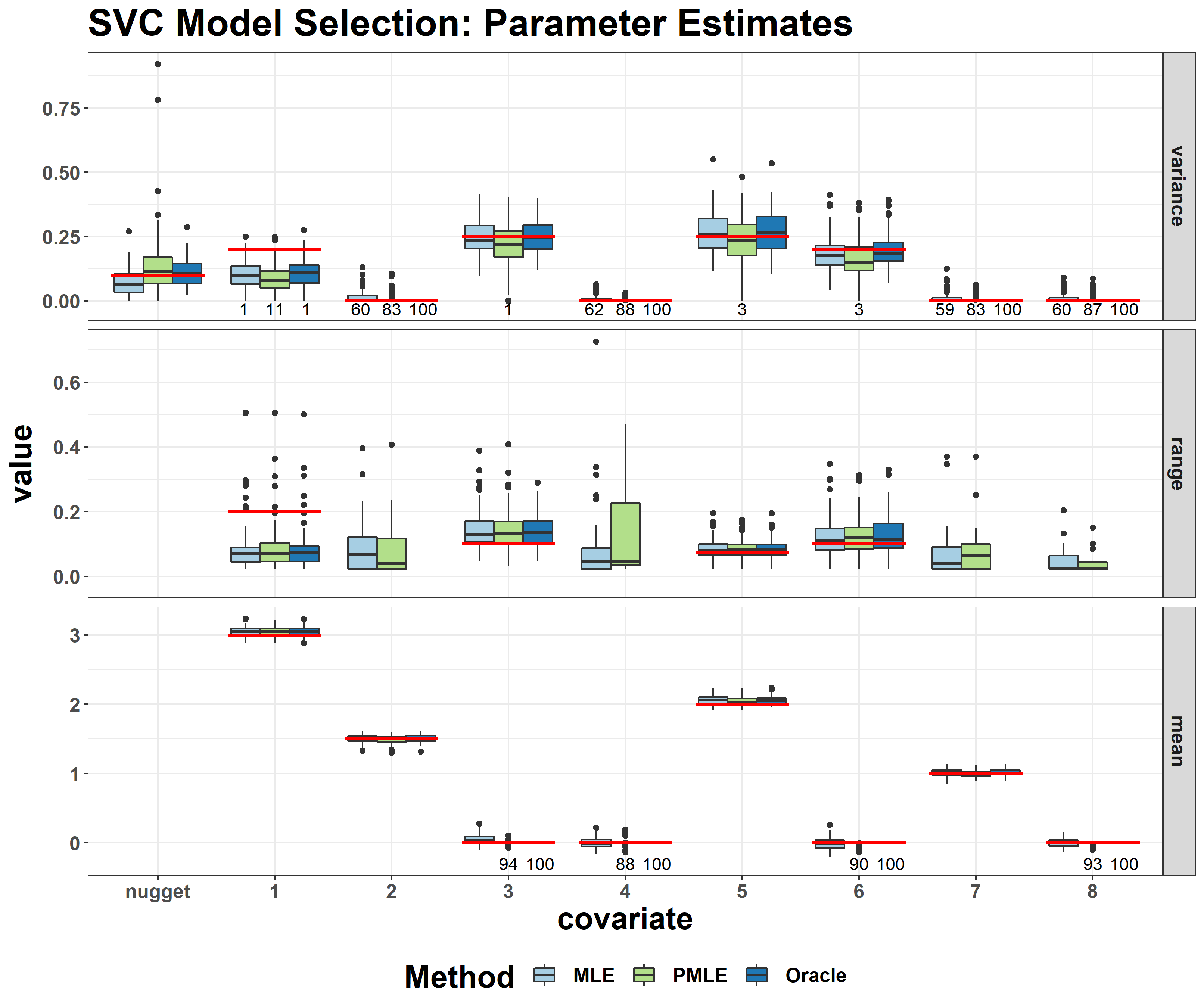}
\caption{Parameter estimates as box plots and, if any, number of zero estimates. Simulation setup: $N = 100$ simulation runs, $n = 225$ observations on a perturbed $15\times 15$ grid. Red line are the true values, c.f. \eqref{eq:true_mu} and Table~\ref{tab:simcovpars}.}\label{fig:par_est}
\end{center}
\end{figure}

Second, we turn to the actual parameter estimates. They are depicted in Figure~\ref{fig:par_est} and grouped by variances, ranges, and means. The true values are indicated by red lines. Further, the number of zero estimates is given, if there are any. As one can see, the \textsf{PMLE} returns sparse estimates in the fixed effects as well as the variances. We can clearly observe that for all methods, the estimation of the mean parameters is much more precise compared to the covariance parameters. However, the MLE method is not able to return mean parameters identical to zero. For the variances, we observe that both the \textsf{MLE} and \textsf{PMLE} are capable of obtaining sparse estimates. Here, the additional penalization of the variances manifests itself in the higher number of zero estimates. On average, we observe an increase of 25 correctly zero estimated variance parameters. The penalty has a minor downside as there are some zero estimates, where the true variance is unequal zero. In that case, the nugget compensates for unattributed variation by the covariates which manifests itself in large variance estimates, c.f.~outliers of box plot for estimated nugget variance, Figure~\ref{fig:par_est}.

Finally, we summarize the results of the simulation study in Table~\ref{tab:sumsim}. For each simulation and methodology $m$, we compute the relative model error (RME) which is defined as
\begin{align*}
	\textnormal{RME}(m) =  \frac{ \| \y - \hat{\y}(m) \|_1}{ \| \y - \bar{y} \cdot \1_n \|_1}.
\end{align*}
The median relative model error (MRME), i.e., the median over all RME, is provided in Table~\ref{tab:sumsim} and is smallest for \textsf{MLE}. This comes at no surprise given the high degree of flexibility of an SVC model. It shows that some kind of variable selection is desirable in order to counter over-fitting. The MRME for \textsf{PMLE} is similar to the one of \textsf{Oracle}. It is not identical as the parameter estimates cannot fully mimic the behavior of the oracle. This is summarized in the second part of Table~\ref{tab:sumsim}, where the number of estimated zero parameters are given both within the mean effects of $\bmu$ and the random effects of $\btheta$. The average number of correctly identified zero parameter (C) and incorrectly identified zero parameters (IC) over all simulations is provided. \textsf{PMLE} introduces sparse estimates for the fixed effects while there are no IC in the fixed effects. Further, \textsf{PMLE} substantially increases the C for the random effects. The downside is a slight increase in the IC. It is also worth mentioning that both the \textsf{MLE} and \textsf{PMLE} do not incorrectly estimate any mean parameter as zero. 

\begin{table}[ht]
\centering
\caption{Median relative model error (MRME) and average number of correctly (C) and incorrectly (IC) estimated zero parameters, divided into fixed effects and random effects, i.e., the SVCs.} 
\label{tab:sumsim}
\begin{tabular}{lrrrrr}
  \hline \textbf{Method} & \textbf{MRME} & \multicolumn{2}{c}{\textbf{Fixed Effects}} & \multicolumn{2}{c}{\textbf{Random Effects}} \\  \cmidrule(lr){3-4}  \cmidrule(lr){5-6}$m$ &  & C & IC & C & IC \\ 
  \hline
\textsf{PMLE} & 0.035 & 3.65 & 0.00 & 3.41 & 0.18 \\ 
  \textsf{MLE} & 0.019 & 0.00 & 0.00 & 2.41 & 0.01 \\ 
  \textsf{Oracle} & 0.032 & 4.00 & 0.00 & 4.00 & 0.01 \\ 
   \hline
\end{tabular}
\end{table}

\subsection{Discussion}

In this section, we summarize our results of the simulation study. Our first focus is the selection of covariance parameters, where the overall performance is not on par with ones of the mean parameters. However, this comes at no surprise as covariance parameters are known to be more difficult to estimate than mean parameters. Further, we can observe a familiar behavior from variable selection of classical linear models. Increasing the sparsity of the model, i.e., having none or only a few SVC, will increase the variance of the nugget effect. 

Overall, our newly suggested PMLE method correctly identifies covariates with no effect in over 90\% of the fixed and 85\% of the random effects. This is notable considering the only drawback is that less than 5\% of covariates were incorrectly estimated to have no effect. 


\section{Application}\label{sec:real}

\subsection{Data}

\begin{figure}
	\centering
	\includegraphics[scale=1]{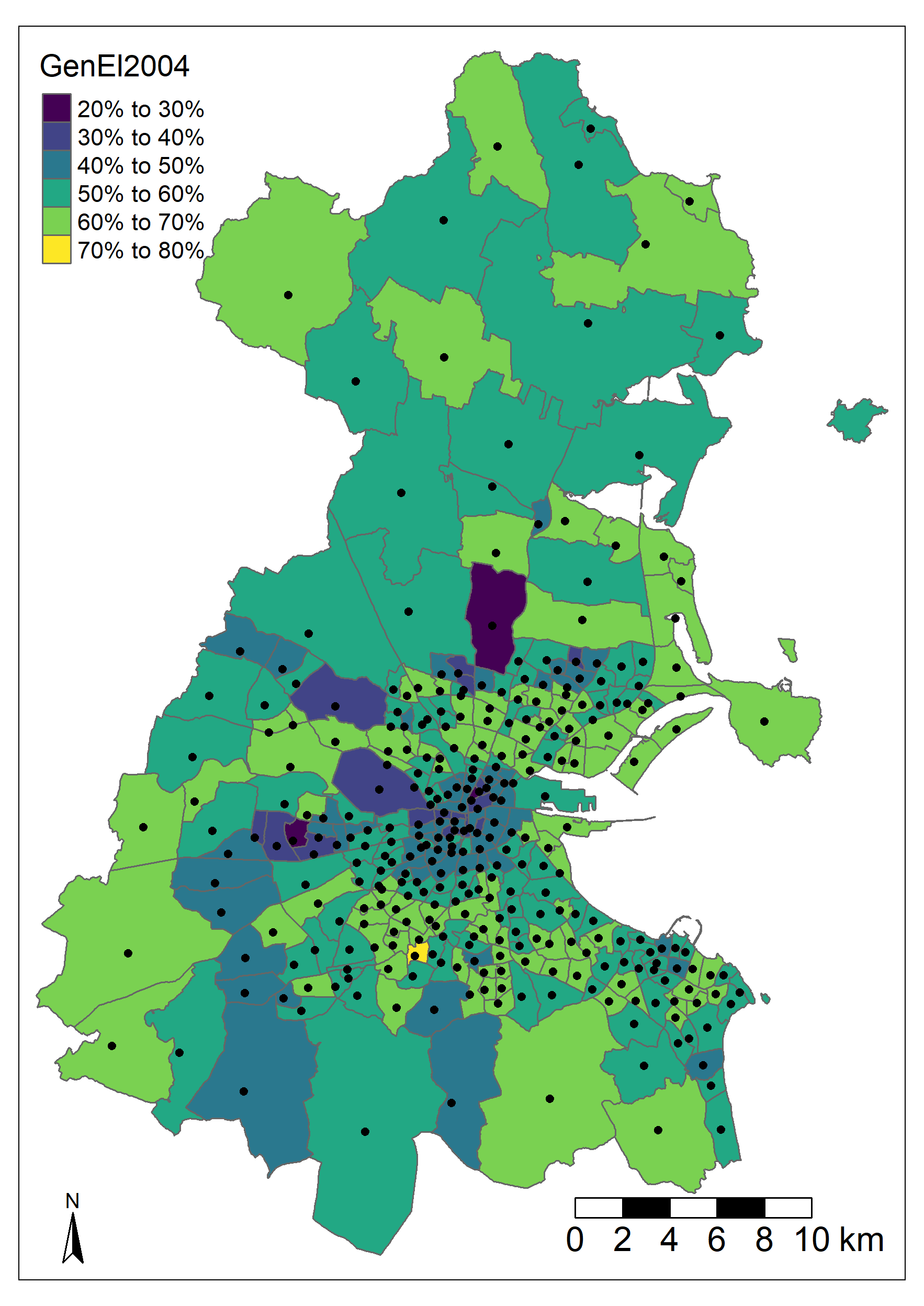}
	\caption{Voter turnout in the 2002 General Elections per ED. The black dots represent the observation locations per ED used in our analysis. They are provided in the data set, too.}\label{fig:01_GenEl2004}
\end{figure}

As an application, we consider the Dublin Voter data set. It consists of the voter turnout in the 2002 General Elections (GE) and 8 other demographic covariates for $n = 322$ electoral divisions (ED) in the Greater Dublin area (Ireland), see Figure~\ref{fig:01_GenEl2004}. All nine  variables are given in percentages and we provide an overview in Table~\ref{tab:DubVoter_sumstat}. The data set was first studied by \citet{Kavanagh2004}. The data set is available in the \textsf{R} package \pkg{GWmodel}. The article by \citet{R:GWmodel} showcases further analysis using the GWR framework. In particular, a GWR selection based on a corrected AIC \citep{Hurvich1998} is conducted. The goal of this section is to do variable selection using the PMLE. Further, we compare our results to linear model-based LASSO and an GWR-based selections.

\begin{table}[ht]
\centering
\caption{Description and summary statistics of Dublin Voter data. The descriptions are taken from the package \pkg{GWmodel}. The summary statistics include the minimum, mean, standard deviation (SD), and maximum. Note that due to an naming error the response of interest is called \texttt{GenEl2004}, although it actually refers to the 2002 GE.} 
\label{tab:DubVoter_sumstat}
\begin{tabular}{llrrrr}
  \hline \textbf{Variable}&	\textbf{Description}& \multicolumn{4}{c}{\textbf{Summary Statistics} [in \%]} \\
	  		  \cmidrule(lr){3-6} & Percentage of population in each ED... & Min. & Mean & SD & Max. \\ 
  \hline
\texttt{DiffAdd} & ...who are one-year migrants & 1.90 & 9.86 & 6.13 & 34.74 \\ 
  \texttt{LARent} & ...who are local authority renters & 0.00 & 15.17 & 24.69 & 100.00 \\ 
  \texttt{SC1} & ...who are social class one (high) & 0.22 & 8.05 & 6.14 & 25.63 \\ 
  \texttt{Unempl} & ...who are unemployed & 1.26 & 7.56 & 5.28 & 31.14 \\ 
  \texttt{LowEduc} & ...who are with little formal education & 0.00 & 0.49 & 0.73 & 9.24 \\ 
  \texttt{Age18\_24} & ...who are age group & 0.14 & 13.43 & 6.06 & 56.55 \\ 
  \texttt{Age25\_44} & ...who are age group & 17.63 & 31.65 & 6.74 & 56.40 \\ 
  \texttt{Age45\_64} & ...who are age group & 7.12 & 20.87 & 5.12 & 34.01 \\ 
  \texttt{GenEl2004} & ...who voted in 2002 GE & 27.98 & 55.61 & 8.71 & 72.91 \\ 
   \hline
\end{tabular}
\end{table}

As one can see in Table~\ref{tab:DubVoter_sumstat}, the summary statistics of the variables vary a lot, which can be an issue in numeric optimization. Therefore, without loss of generality and interpretability, we standardize the data by subtracting the empirical mean and scaling by the empirical standard deviation. We annotate the standardized variables by a prefix ``\texttt{Z.}''.

Further, the observation locations represent the electoral divisions as depicted in Figure~\ref{fig:01_GenEl2004} and are provided as Easting and Northing in meters. We transform them to kilometers to ensure computational stability while staying interpretable with respect to the range parameter. 

\subsection{Models and Methodologies}\label{sec:app_model}

We use two models in our comparison. First, a simple linear regression using the adaptive LASSO for variable selection is defined as
\begin{equation*}
\begin{aligned}
	y_i &= \texttt{Z.GenEl2004}_i \\
	&=\quad \phantom{\mu_1}
	&\phantom{+}\ &\mu_2 \texttt{Z.DiffAdd}_i 
	&+\ &\mu_3 \texttt{Z.LARent}_i \\
	&\quad+ \mu_4 \texttt{Z.SC1}_i 
	&+\ &\mu_5 \texttt{Z.Unempl}_i 
	&+\ &\mu_6 \texttt{Z.LowEduc}_i \\
	&\quad+ \mu_7 \texttt{Z.Age18\_24}_i 
	&+\ &\mu_8 \texttt{Z.Age25\_44}_i 
	&+\ &\mu_9 \texttt{Z.Age45\_64}_i + \varepsilon_i.
\end{aligned}
\end{equation*}

Since we use the standardized covariate $\texttt{Z.GenEl2004}$ as our response, we do not expect an intercept in the model. The coefficients are denoted as $\mu_j$ in accordance with our previous notation, i.e., these are only mean effects, and \textsf{ALASSO} denotes the adaptive LASSO used to estimate the mean effects. It is implemented using the \textsf{R} package \pkg{glmnet} \citep{Friedman2010}, where the corresponding adaptive weights are given by an ordinary least squares estimate and the shrinkage parameter is estimated by cross-validation. The second model is a full SVC model including an intercept:
\begin{equation*}
\begin{aligned}
	y_i &= \texttt{Z.GenEl2004}_i \\
	&=\quad \beta_1(s_i) 
	&+\ &\beta_2(s_i) \texttt{Z.DiffAdd}_i 
	&+\ &\beta_3(s_i) \texttt{Z.LARent}_i \\
	&\quad+ \beta_4(s_i) \texttt{Z.SC1}_i 
	&+\ &\beta_5(s_i) \texttt{Z.Unempl}_i 
	&+\ &\beta_6(s_i) \texttt{Z.LowEduc}_i \\
	&\quad+ \beta_7(s_i) \texttt{Z.Age18\_24}_i 
	&+\ &\beta_8(s_i) \texttt{Z.Age25\_44}_i 
	&+\ &\beta_9(s_i) \texttt{Z.Age45\_64}_i + \varepsilon_i.
\end{aligned}
\end{equation*}
The SVCs have yet to be specified according to the method being used to estimate the model, but generally, they do contain mean effects, i.e., the SVCs are not centered around zero. As for the intercept, we expect a zero mean effect. However, there might be some spatially local structures that can be captured using an SVC.

In the case of the GP-based SVC model as given in \eqref{eq:SVCmodelsingle}, the $\beta_j(s_i)$ can simply be partitioned into the mean effect $\mu_j$ and the random effect $\eta_j(s_i)$. For the GWR, $\beta_j(s_i)$ is defined as the generalized least square estimate where observations are geographically weighted. As a kernel to weight the observations based on their distances, we are using an exponential function. The kernel relies on an adaptive bandwidth that is selected using a corrected AIC \citep{Hurvich1998}.

We use MLE as well as PMLE to estimate the full SVC model, again labeled \textsf{MLE} and \textsf{PMLE}, respectively, and implemented using \pkg{varycoef}. For further reference, we also report the results of an GWR-based variable selection labeled \textsf{GWR} and using the \textsf{R} package \pkg{GWmodel} as well as an adaptive LASSO labeled \textsf{ALASSO} and using the \textsf{R} package \pkg{glmnet}.

\subsection{Results}\label{sec:DubVoter_results}

In this section, we present the results for all model-based approaches, i.e., \textsf{ALASSO}, \textsf{MLE}, and \textsf{PMLE}, as not all comparison measures of \textsf{GWR} are defined. In terms of variable selection, we provide the estimated shrinkage parameters, number of non-zero estimates of the mean and the variances, the log likelihood $\ell$ and the BIC in Table~\ref{tab:DubVoter_BIC}. As one can see, the \textsf{ALASSO} gives a very sparse model with a relative small log likelihood. As expected, \textsf{MLE} does not provide any sparse fixed effects but selects only 6 of the potentially 9 random effects. With \textsf{PMLE} we obtain a sparser model. Over all, the smallest BIC is achieved by \textsf{PMLE} followed by \textsf{MLE} and \textsf{ALASSO}.
 
\begin{table}[ht]
\centering
\caption{Overview of estimated shrinkage parameters (Shrink.\ Par.), model complexity (MC), goodness of fit (GoF), and combined information criterion (IC) of all model-based approaches.} 
\label{tab:DubVoter_BIC}
\begin{tabular}{lrrrrrr}
  \hline \textbf{Method} &	 \multicolumn{2}{c}{\textbf{Shrink.\ Par.}} &  \multicolumn{2}{c}{\textbf{MC}} & \textbf{GoF} & \textbf{IC} \\ \cmidrule(lr){2-3}   \cmidrule(lr){4-5} $m$ & $\hat{\lambda}_\bmu$ & $\hat{\lambda}_\btheta$ & $\|\hat{\bmu}\|_0$ & $\|\hat{\bsigma}^2\|_0$ & $\ell$ & BIC \\ 
  \hline
\textsf{ALASSO} & 0.49 & $-$ & 3 & $-$ & $-$303.9 & 625.2 \\ 
  \textsf{MLE} & $-$ & $-$ & 9 & 6 & $-$264.0 & 614.7 \\ 
  \textsf{PMLE} & 0.15 & $9.1 \times 10^{-6}$ & 7 & 5 & $-$264.3 & 563.3 \\ 
   \hline
\end{tabular}
\end{table}

In Table~\ref{tab:DubVoter_parest}, the estimated parameters are provided. For the fixed effects, we observe that \textsf{PMLE} results in zero estimates for the intercept and \texttt{Z.LowEduc}. These fixed effects are only a subset of what \textsf{ALASSO} estimated to be zero.

\begin{table}[ht]
\centering
\caption{Parameter estimates of model-based approaches. We use {``--"} for not available values. To ease readability, estimates identical to 0 are given by a sole zero, i.e., \emph{not} by ``0.000". Range estimates, where the corresponding variance was estimated to be equal to zero, are given in italic font as they cannot be interpreted.} 
\label{tab:DubVoter_parest}
\begin{tabular}{llrrrrrrr}
  \hline \textbf{Variable}&	$j,k$ &  \multicolumn{3}{c}{\textbf{Mean} $\hat{\mu}_j$} & \multicolumn{2}{c}{\textbf{Range} $\hat{\rho}_k$} & \multicolumn{2}{c}{\textbf{Variance} $\hat{\sigma}^2_k$} \\ \cmidrule(lr){3-5}  \cmidrule(lr){6-7}  \cmidrule(lr){8-9}  &  & \textsf{ALASSO} & \textsf{MLE} & \textsf{PMLE} & \textsf{MLE} & \textsf{PMLE} & \textsf{MLE} & \textsf{PMLE} \\ 
  \hline
Intercept & 1 & --\phantom{.000} & $-$0.020 & 0\phantom{.000} & 2.780 & 2.789 & 0.102 & 0.101 \\ 
  \texttt{Z.DiffAdd} & 2 & 0\phantom{.000} & $-$0.084 & $-$0.039 & 1.703 & 1.968 & 0.075 & 0.084 \\ 
  \texttt{Z.LARent} & 3 & $-$0.224 & $-$0.233 & $-$0.222 & \emph{4.627} & \emph{4.627} & 0\phantom{.000} & 0\phantom{.000} \\ 
  \texttt{Z.SC1} & 4 & 0\phantom{.000} &  0.158 &  0.119 & 4.783 & \emph{4.783} & 0.006 & 0\phantom{.000} \\ 
  \texttt{Z.Unempl} & 5 & $-$0.450 & $-$0.503 & $-$0.509 & 3.293 & 3.275 & 0.019 & 0.018 \\ 
  \texttt{Z.LowEduc} & 6 & 0\phantom{.000} &  0.001 & 0\phantom{.000} & \emph{3.794} & \emph{3.794} & 0\phantom{.000} & 0\phantom{.000} \\ 
  \texttt{Z.Age18\_24} & 7 & 0\phantom{.000} & $-$0.072 & $-$0.055 & \emph{4.865} & \emph{4.865} & 0\phantom{.000} & 0\phantom{.000} \\ 
  \texttt{Z.Age25\_44} & 8 & $-$0.209 & $-$0.244 & $-$0.222 & 3.342 & 3.408 & 0.056 & 0.059 \\ 
  \texttt{Z.Age45\_64} & 9 & 0\phantom{.000} & $-$0.107 & $-$0.070 & 6.297 & 6.303 & 0.029 & 0.030 \\ 
   \hline
\end{tabular}
\end{table}

In the covariance effects, we observe an interesting behavior. We note that \textsf{MLE} gives zero estimates for one third of the SVC variances even without any penalization. This coincides with previous results from the simulation study. As expected, the non-zero variance covariates of \textsf{PMLE} are a subset of the ones of \textsf{MLE}. Further details on the CD algorithm for \textsf{PMLE} are given in Appendix~\ref{app:CD_iter}.

Overall, the PMLE method results in less zero fixed effects compared to the adaptive Lasso. There is only one variable which does not enter the model at all, namely  \texttt{Z.LowEduc}, and only one other that does have a zero mean effect, namely the intercept. However, the linear model does not take the spatial structures into account. Considering the spatial structure apparently triggers the inclusion of additional fixed effects. The model gains complexity but we have to consider that this might be necessary in order to account for the spatial structures within the data. Therefore, SVC model selection remains a trade-off between model complexity and goodness of fit which we will address in the following section using cross-validation.

\subsection{Cross-Validation}

In the last section we examined the goodness of fit combined with model complexity as well as parameter estimation. We now turn to predictive performance. Here, we expect that due to high degree of flexibility of full SVC models, methods without any kind of variable selection could result in over-fitting on the training data in connection with biased spatial extrapolation on the prediction data set.

\begin{figure}
	\centering
	\includegraphics[width=\textwidth]{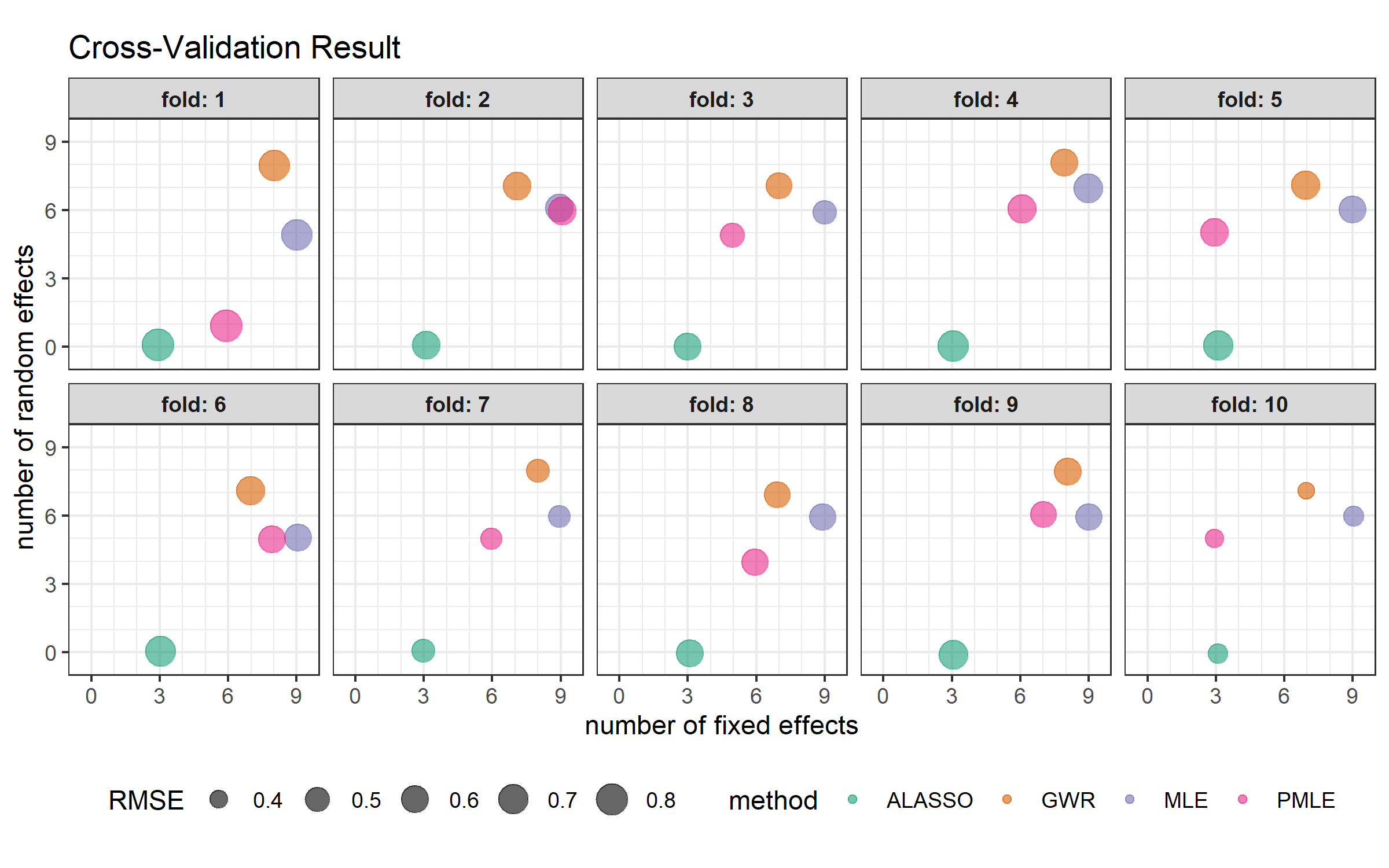}
	\caption{Bubble plot visualizing the results of the cross-validation. The location of the bubbles corresponds to number of fixed and random effects selected by model. Note that the locations are slightly jittered to ensure overlapping methods are still visible. The size of the bubbles corresponds to the out-of-sample RMSE. The facets show individual results per fold.}\label{fig:CV_bubble}
\end{figure}

To examine this, we conduct a classical 10-fold cross-validation. That is, we randomly divide the Dublin Voter data set into ten folds of size 32 or 33, i.e., each observation $i$ falls into one of the following disjoint sets $\mathcal{S}_1, ..., \mathcal{S}_{10}: \bigcup_{\iota = 1}^{10} \mathcal{S}_\iota = \{ 1, ..., 322\}$. For each $\iota = 1, ..., 10$, the variable selection and model fitting of all four methods $m \in \{ \textsf{ALASSO}, \textsf{GWR}, \textsf{MLE}, \textsf{PMLE} \}$ is conducted on the union of nine sets (short hand notation $\mathcal{S}_{-\iota}$) with one set left out for validation ($\mathcal{S}_\iota$). We report the out-of-sample rooted mean squared errors (RMSE) on $\mathcal{S}_\iota$ for each method $m$ denoted $\textnormal{RMSE}_\iota(m)$. This procedure is repeated for each $\iota$. The results are visualized in Figure~\ref{fig:CV_bubble} and Table~\ref{tab:RMSE_summary}. 

First, we observe that the size of the bubbles, i.e., the RMSE, varies more between different folds than between different methods. To analyze the differences in predictive performance, we provide the mean and standard deviation per method in Table~\ref{tab:RMSE_summary}. While the predictive performance of the estimated SVC models are similar, \textsf{ALASSO} falls behind. 

\begin{table}[ht]
\centering
\caption{Mean and standard deviation (SD) of RMSE per method.} 
\label{tab:RMSE_summary}
\begin{tabular}{lrr}
  \hline \textbf{Method} &	\multicolumn{2}{c}{\textbf{RMSE}} \\
	  		  \cmidrule(lr){2-3}$m$ & Mean & SD \\ 
  \hline
\textsf{PMLE} & 0.580 & 0.114 \\ 
  \textsf{MLE} & 0.579 & 0.103 \\ 
  \textsf{GWR} & 0.589 & 0.101 \\ 
  \textsf{ALASSO} & 0.644 & 0.128 \\ 
   \hline
\end{tabular}
\end{table}

Second, we address the number of selected fixed and random effects. Upon closer inspection of Figure~\ref{fig:CV_bubble}, we see notable variations of the number of selected fixed and random effects over different folds for all methods but \textsf{ALASSO}. We attribute this to the relatively small sample size. As already observed in Section~\ref{sec:DubVoter_results}, the number of selected mean effects is higher for selection methods PMLE and GWR on full SVC models than the adaptive LASSO on a linear model. However, the inclusion of more fixed effects and accounting the spatial structure using SVCs clearly surpasses the predictive performance of a simple linear model.

\section{Conclusion}\label{sec:conc}

We present a novel methodology to efficiently select covariates in Gaussian process-based SVC models. We propose to perform variable selection for both fixed effects and spatially varying coefficients by respectively penalizing the fixed regression coefficients and the variance of the spatial random coefficients using $L_1$ penalties. Further, we show how optimization of the resulting objective function can be done using a coordinate descent algorithm. The latter can be done efficiently by relying on existing efficient methodology for the fixed effects part and by parallelizing the optimization procedure.  

In a simulation study, we show that our method is capable of spatial variance component as well as fixed effect selection. In real data application, we find that our novel method results in the best in-sample fit with the lowest BIC among the considered approaches and gives high predictive accuracy when doing cross-validation. While achieving a similar predictive performance as the unpenalized MLE, the PML estimated models are sparser and have lesser model complexity. Examining the influence of possibly new covariates in SVC models, we recommend our newly proposed PMLE over a classical MLE for variable selection. 

Our novel variable selection methodology has been implemented specifically with SVC models in mind. However, with some adaptions, one can apply it to generalized linear models that fit in our underlying model assumptions. Further future work on our PMLE can include individual weighting of fixed and random effects via the BIC. One could add weights to the specific model complexity penalization as, usually, the addition of another fixed effect to an SVC model is not equal to adding another random effect in terms of, say, computational work load or model complexity.

\section*{Declarations}

\subsection*{Acknowledgment}

JD sincerely thanks Lucas Kook for the valuable exchange and help with respect to the MBO implementation. 

\subsection*{Funding}

JD and FS gratefully acknowledge the support of the \emph{Swiss Agency for Innovation} (\emph{innosuisse} project number 28408.1 PFES-ES). RF gratefully acknowledges the support of the \emph{Swiss National Science Foundation} (SNSF grant 175529).

\subsection*{Conflicts of Interest}

The authors declare no competing interests. 

\subsection*{Availability of Data and Material}

The data from the simulation is available online under \url{https://git.math.uzh.ch/jdambo/open-access-svc-selection-paper}. The data used in the application is given in the \textsf{R} package \pkg{GWmodel} \citep{R:GWmodel}.

\subsection*{Code Availability}

The code for the simulation and application is available online under \url{https://git.math.uzh.ch/jdambo/open-access-svc-selection-paper}. Our novel proposed method is implemented in the \textsf{R} package \pkg{varycoef} \citep{R:varycoef}.

\bibliographystyle{../../../my_bib/chicago}

\bibliography{../../../my_bib/mybib}


\newpage
\pagenumbering{arabic}
\renewcommand*{\thepage}{A\arabic{page}}
\appendix

\section{Appendix to: \protect Joint Variable Selection of both Fixed and Random Effects for Gaussian Process-based Spatially Varying Coefficient Models}

\subsection{Proofs}

\begin{proof}[Proof of Proposition~\ref{prop:gr0}]\label{proof:gr0}
	Let $U_A \in \mathbb{R}^{n\times n}$ be the distance matrix under some anisotropic geometric norm defined by a positive-definite matrix $A\in \mathbb{R}^{d\times d}$, i.e., $\left(U_A \right)_{lm} = u_{lm} := \|s_l - s_m\|_A$. Let $r: \mathbb{R}_{\geq 0} \rightarrow \left[0, 1\right]$ be a correlation function where by $r(U_A)$ we denote the component-wise evaluation, i.e., $\left(r(U_A)\right)_{lm} = r(u_{lm})$. We have:
	\begin{align*}
		\frac{\partial}{\partial \rho_\kappa} \bSigma_\Y(\btheta)  &= \frac{\partial}{\partial \rho_\kappa}\left( \sum_{k = 1}^q \left(\w^{(k)}{\w^{(k)}}^\top\right) \odot \bSigma_k + \tau^2 \I_{n\times n} \right) \\
			&= \left(\w^{(k)}{\w^{(k)}}^\top\right) \odot \frac{\partial}{\partial \rho_\kappa} \bSigma_\kappa = \\
			&= \left(\w^{(k)}{\w^{(k)}}^\top\right) \odot \sigma_\kappa^2 r'\left(\frac{U_A}{\rho_\kappa}\right)\odot\left(-\frac{U_A}{\rho_\kappa^2}\right) .
	\end{align*}
	With $|r'|<C$, we have $\frac{\partial}{\partial \rho_\kappa} \bSigma_\Y(\b_\kappa) = \0_{n\times n}$ and recall that $\bSigma_\Y(\b_\kappa)$ is well-defined and invertible. With the identities $(\log \det M)' = \textnormal{tr} (M^{-1} M')$ and $\left(M^{-1}\right)' = -M^{-1} M' M^{-1}$ for some quadratic matrix $M$, we obtain 
\begin{align*}
	\frac{\partial}{\partial \rho_\kappa} f (\btheta) & = \textnormal{tr} \left(  {\bSigma_\Y(\btheta)}^{-1} \frac{\partial}{\partial \rho_\kappa} \bSigma_\Y(\btheta) \right)  \\ &+ (\y-\X\bmu^{(t+1)})^\top \left( - {\bSigma_\Y(\btheta) }^{-1} \left(\frac{\partial}{\partial \rho_\kappa} \bSigma_\Y(\btheta) \right) {\bSigma_\Y(\btheta) }^{-1} \right) (\y-\X\bmu^{(t+1)})
\end{align*}	
and therefore $\frac{\partial}{\partial \rho_\kappa} f(\b_\kappa) = 0$.
	
\end{proof}

\subsection{Perturbed Grid}\label{app:per_grid}

\begin{figure}
\begin{center}
\includegraphics[width = 0.9\textwidth]{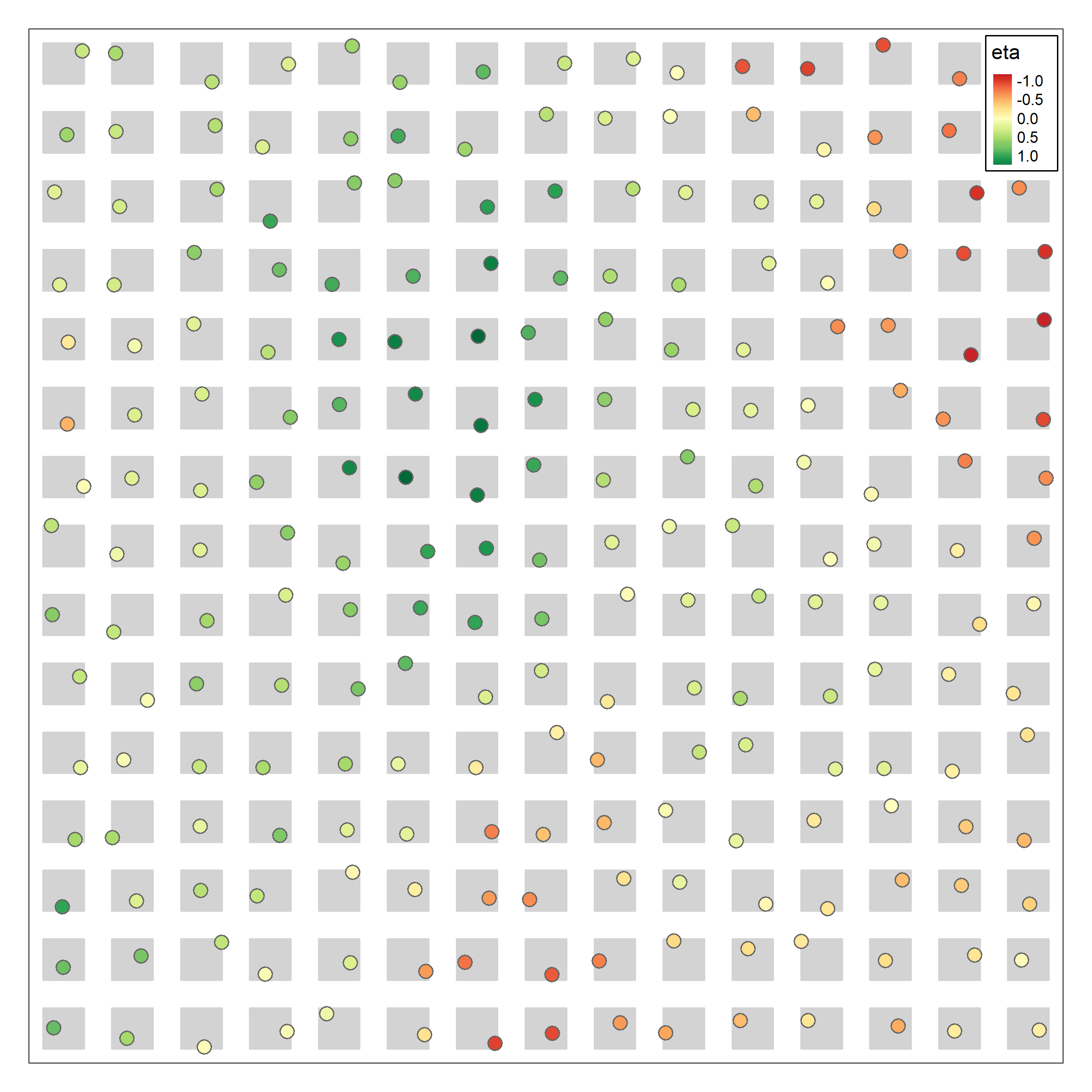}
\caption{Perturbed grid of size $15 \times 15$ within the unit square (black border). The grey squares indicate the sampling domian. The colored points are the observation locations with corresponding value of the SVC, i.e., $\bfeta(\s)$.}\label{fig:per_grid}
\end{center}
\end{figure}

A perturbed grid is used to sample the observation locations. It consists of $15 \times 15$ sampling domains arranged as a regular grid. Each sampling domain is a square surrounded by a thin margin. We sample an observation location uniformly from each square in the sampling domains. The true effect of the SVC is then sampled from an GP at corresponding observation locations. In Figure~\ref{fig:per_grid} such an example is provided. This is repeated for all $N = 100$ simulation runs.

\subsection{Numeric Optimization}\label{app:num_opt}

We provide further details for the numeric optimization. In all ML and PML methods, we use the \R\ package \pkg{optimParallel} \citep{R:optimParallel}. In the simulation study, recall that we are on a $m \times m$ perturbed grid, where we have $m = 15$ observations along one side. Here, we set $\sigma_k^2 \geq 0, \rho_k \geq (3m)^{-1}$, and $\tau^2 \geq 10^{-4}$ as lower bounds. The lower bounds in the case of the ranges are motivated by the \emph{effective range} of an exponential GP. It is $3\rho$, where $\rho$ is the corresponding range of the GP. Since the smallest distances between neighbors on a perturbed grid are $1/m$ on average, it is not feasible to model GPs with ranges smaller than a third of that. This prevents individual SVCs to take on the role of a nugget effect.

\subsection{Coordinate Descent Iterations}\label{app:CD_iter}

\begin{figure}
	\centering
	\includegraphics[width=\textwidth]{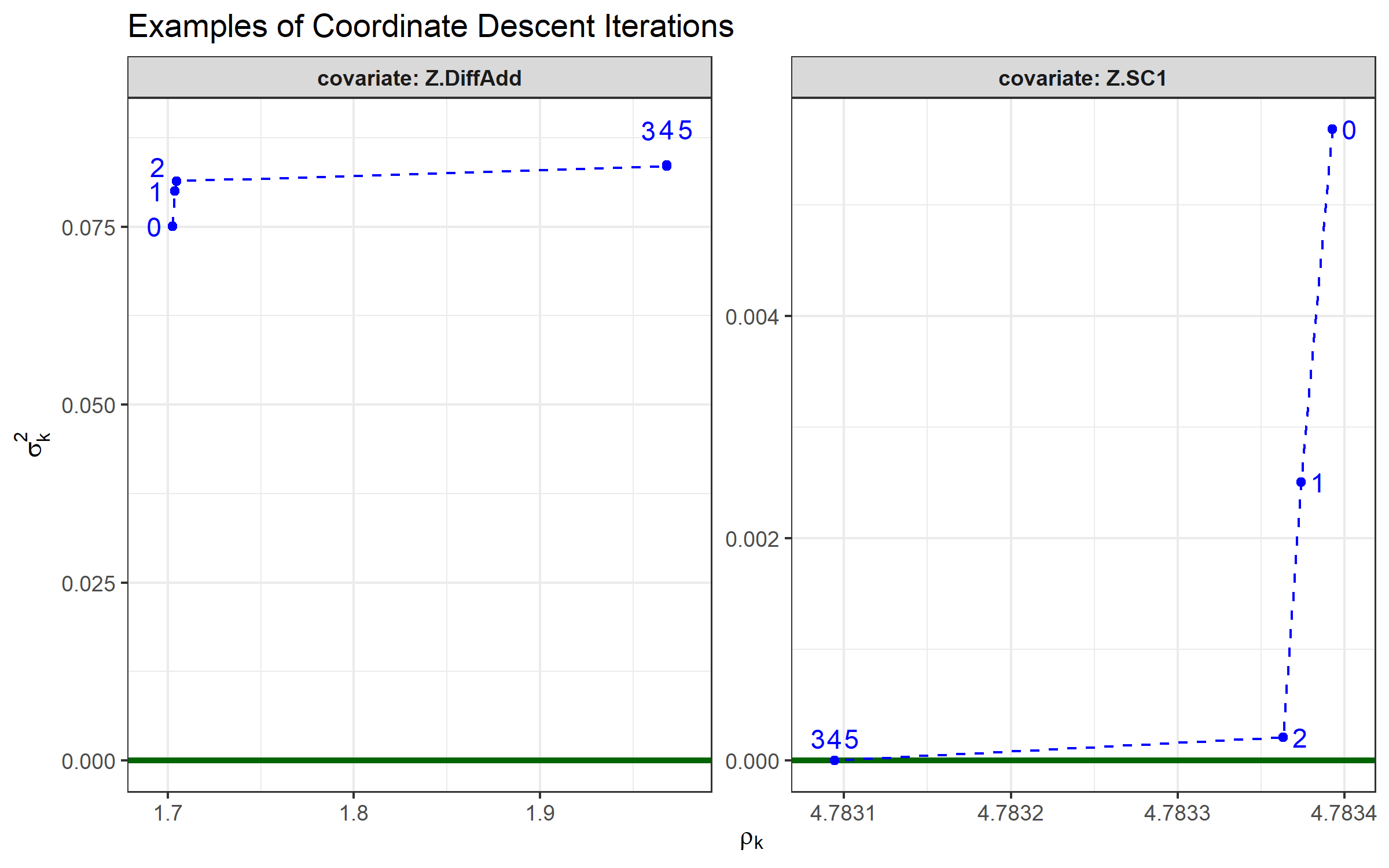}
	\caption{Covariance parameters trajectories of two SVCs for covariates \texttt{Z.DiffAdd} and \texttt{Z.SC1} in coordinate descent.}\label{fig:CD_iter}
\end{figure}

We briefly discuss the coordinate descent in \textsf{PMLE} from Sections~\ref{sec:app_model} and \ref{sec:DubVoter_results}. For the two covariates \texttt{Z.DiffAdd} and \texttt{Z.SC1}, we give the covariance parameters for respective SVCs at individual steps $t$ in the coordinate descent in Figure~\ref{fig:CD_iter}. We have chosen these two covariates as both their variances have the largest absolute difference between the initial and final values. Additionally, one of the variances shrunk to 0 where the other did not. In total, we have $T = 5$ iterations of the CD algorithm and the individual covariance parameters $(\rho_k^{(t)}, {\sigma_k^2}^{(t)})$ for $t = 0, ..., T$ and $k = 2, 4$ are given by the blue dots with annotated iteration step $t$. Note that for both covariates the covariance parameters of the last 3 iteration steps are (almost) identical. The lower bound of the variance is indicated by the dark green line. Figure~\ref{fig:CD_iter} clearly visualizes the effect of adding the penalties on the variance parameters as both optimization trajectories first evolve along the variance axis, before the range is adjusted further on.    For $\rho_4^{(t)}, t \geq 3$, we see that no further adjustments are made beside the fact that $\rho_4$ is unidentifiable under $\sigma_4^2 = 0$ (Proposition~\ref{prop:gr0}).

\end{document}